\documentstyle[aps,prl,preprint]{revtex} 

\begin{document}

\title{Nonlinear optics with less than one photon}
\author{K.J. Resch, J.S. Lundeen, and A.M. Steinberg}
\address{Department of Physics, University of Toronto,\\
60 St. George Street, Toronto, Ontario, Canada\ M5S 1A7}
\maketitle

\begin{abstract}
We demonstrate suppression and enhancement of spontaneous parametric
down-conversion via quantum interference with two weak fields from a local
oscillator (LO). \ Pairs of LO\ photons are observed to upconvert with high
efficiency for appropriate phase settings, exhibiting an effective
nonlinearity enhanced by at least 10 orders of magnitude. \ This constitutes
a two-photon switch, and promises to be useful for a variety of nonlinear
optical effects at the quantum level.
\end{abstract}

\newpage \tightenlines

Nonlinear effects in optics are typically limited to the high-intensity
regime, due to the weak nonlinear response of even the best materials. \ An
important exception occurs for resonantly enhanced nonlinearities, but these
are restricted to narrow bandwidths. \ Nonlinear effects which are
significant in the low-photon-number regime would open the door to a field
of \emph{quantum} nonlinear optics. \ This could lead to optical switches
effective at the two-photon level (i.e. all-optical quantum logic gates);
quantum solitons (e.g. two-photon bound states \cite{ray and ivan}); and a
host of other phenomena. \ In this experiment, we demonstrate an effective
two-photon nonlinearity mediated by a strong classical field. \ Quantum
logic operations have already been performed in certain systems including
trapped ions \cite{wineland}, NMR \cite{NMR quant. comp}, and cavity QED %
\cite{cavity guys}, but there may be advantages to performing such
operations in an all-optical scheme -- including scalability and relatively
low decoherence. \ A few schemes have been proposed for producing the
enormous nonlinear optical responses necessary to perform quantum logic at
the single-photon level. \ Such schemes involve coherent atomic effects
(slow light \cite{slow light} and E.I.T. \cite{eit}) or photon-exchange
interactions \cite{photon exchange interactions}. \ We recently demonstrated
that photodetection exhibits a strong two-photon nonlinearity \cite{resch},
but this is not a coherent response, as it is connected to the amplification
stage of measurement. \ While there has been considerable progress in these
areas, coherent nonlinear optical effects have not yet been observed at the
single-photon level. \ In a typical setup for second-harmonic generation,
for instance a peak intensity on the order of 1 GW/cm$^{2}$ is required to
provide an upconversion efficiency on the order of 10\%. \ In the experiment
we describe here, beams with peak intensities on the order of 1 mW/cm$^{2}$
undergo second-harmonic generation with an efficiency of about 1\% --
roughly 11 orders of magnitude higher than would be expected without any
enhancement.

Our experiment relies on the process of spontaneous parametric
down-conversion. \ If a strong laser beam with a frequency 2$\omega $ passes
through a material with a nonzero second-order susceptibility, $\chi ^{(2)}$%
, then pairs of photons with nearly degenerate frequencies, $\omega $, can
be created. \ In past experiments, interference phenomena have been observed
between weak classical beams and pairs of down-conversion beams \cite%
{rarity, japanese group, walmsley}. \ Although spontaneously down-converted
photons have no well-defined phase (and therefore do not display first-order
interference), the \emph{sum} of the phases of the two beams is fixed by the
phase of the pump. \ Koashi et al. \cite{japanese group} have observed this
phase relationship experimentally using a local oscillator (LO)\
harmonically related to the pump. \ More recently Kuzmich et al. \cite%
{walmsley} have performed homodyne measurements to directly demonstrate the
anticorrelation of the down-converted beams' phases. \ Some proposals for
tests of nonlocality \cite{hardy} have relied on the same sort of effect. \
Such experiments involve beating the down-converted light against a local
oscillator at one or more beam splitters, and hence have multiple output
ports. \ The interference causes the photon-correlations to shift among the
various output ports of the beam splitters. \ 

In contrast, in this experiment the actual photon-pair production rate is
modulated. \ A simplified cartoon schematic of our experiment is shown in
Fig. 1. \ A nonlinear crystal is pumped by a strong classical field,
creating pairs of down-converted photons in two distinct modes (solid
lines). \ Local oscillator beams are superposed on top of the
down-conversion modes through the nonlinear crystal and are shown as dashed
lines. \ A single-photon counting module (SPCM)\ is placed in the path of
each mode. \ To lowest order there are two Feynman paths that can lead to
both detectors firing at the same time (a coincidence event). \ A
coincidence count can occur either from a down-conversion event (Fig. 1b.),
or from a pair of LO\ photons (Fig. 1c.). \ Interference occurs between
these two possible paths provided they are indistinguishable. \ Depending on
the phase difference between these two paths ($\varphi $), we observe
enhancement or suppression of the coincidence rate. \ A phase-dependent rate
of photon-pair production has been observed in a previous experiment using
two pairs of down-converted beams from the same crystal \cite{zeilinger}. \
By contrast, our experiment uses two independent LO fields which can be from
classical or quantum sources, and subject to external control. \ If the
phase between the paths (Fig. 1b, 1c) is chosen such that coincidences are
eliminated, then photon pairs are removed from the LO beams by upconversion
into the pump mode. \ If, however, one of the LO\ beams is blocked, then
those photons that would have been upconverted are now transmitted through
the crystal. \ This constitutes an optical switch in which the presence of
one LO\ field controls the transmission of the other LO field -- even when
there is less than one photon in the crystal at a time. \ This switch does
have certain limitations. \ First, it is inherently noisy due to the
spontaneous down-conversion, which leads to coincidences even if both LO
beams are blocked. \ Second, since the switch relies on interference, and
hence phase, it does not occur between photon pairs but between the \emph{%
amplitudes} to have a photon pair. \ While this may limit the usefulness of
the effect as the basis of a ``photon-transistor,'' a simple extension
should allow it to be used for conditional-phase operations.

\bigskip

In order for the down-conversion beams to interfere with the laser beams,
they must be indistinguishable in all ways (including frequency, time,
spatial mode, and polarization). \ Down-conversion is inherently broadband
and exhibits strong temporal correlations; the LOs must therefore consist of
broadband pulses as well. \ We use a modelocked Titanium:Sapphire (Ti:Sa)
laser operating with a central wavelength of 810 nm (Fig. 2). \ It produces
50-fs pulses at a rate of 80 MHz. \ This produces the LO\ beams, and its
second-harmonic serves as the pump for the downconversion. \ Thus the
downconversion is centered at the same frequency as the LO, and the LOs and
the downconverted beams have similar bandwidths of around 30 nm. \ To
further improve the frequency overlap, we frequency post-select the beams
using a narrow bandpass\ (10-nm) interference filter \cite{sasha}. \ As this
is narrower than the bandwidth of the pump, it erases any frequency
correlations between the downconversion beams. \ This removes the frequency
correlations between the down-conversion beams. \ In addition to spectral
indistinguishability, the two light sources must possess spatial
indistinguishability. \ The down-conversion beams contain strong spatial
correlations between the correlated photon pairs; measurement of a photon in
one beam yields some information about the photon in the other beam. \ Such
information does not exist within a laser beam; since there is only a single
transverse mode, the photons must effectively be in a product state and
exhibit no correlations. \ We therefore select a single spatial mode of the
down-converted light by employing a simple spatial filter. \ The beams are
focused onto a 25-$\mu $m diameter circular pinhole. \ The light that passes
through the pinhole and a 2-mm diameter iris placed 5 cm downstream is
collimated using a 5-cm lens. \ In order to increase the flux of
down-converted photons into this spatial mode, we used a pump focusing
technique related to the one demonstrated by Monken et al. \cite{brazilians}%
. \ The pump laser was focused directly onto the down-conversion crystal. \
Since the coherence area of the down-converted beams is set by the
phase-matching acceptance angle, the smaller pump area reduced the number of
spatial modes being generated at the crystal, improving the efficiency of
selection in a single mode. \ Imaging the small illuminated spot of our
crystal onto the pinhole, we were able to improve the rate of coincidences
after the spatial filter by a factor of 30.

The final condition necessary to obtain interference is to have a
well-defined phase relationship between the LO beams and the down-conversion
beams. \ To achieve this, the same Ti:Sa source laser is split into two
different paths (Fig. 2.). \ The majority of the laser power (90\%)\ is
transmitted through BS1 into path 1, where it is type-I\ frequency-doubled
to produce the strong (approximately 10-mW) classical pump beam with a
central frequency of 405 nm. \ This beam is used to pump our down-conversion
crystal (DC) after the 810-nm fundamental light is removed by colored glass
filters. \ Instead of using down-conversion with spatially separate modes as
shown in Fig. 1, we use type-II down-conversion, in which the photon pairs
are emitted in the same direction but with distinct polarizations. \ The
photon pairs are subsequently spatially filtered, spectrally filtered, and
then split up by the polarizing beam splitter (PBS). \ The
horizontally-polarized photon is transmitted to detector A, and the
vertically-polarized photon is reflected to detector B. \ Detectors A and B
are both single-photon counting modules (EG\&G models SPCM-AQ-131 and
SPCM-AQR-13). \ Path 1 also contains a trombone delay arm which can be
displaced to change the relative phase between paths 1 and 2. \ To create
the LO laser beams, we use the 10\% reflection from BS1 into path 2. \ The
vertically-polarized laser light is attenuated to the single-photon level by
a set of neutral-density (ND)\ filters, and its polarization is then rotated
by 45$^{\circ }$ using a zero-order half-wave plate, so that it serves
simultaneously as LO\ for the horizontal and vertical beams. \ After the
wave-plate, the light passes through a polarizer, which can be used to block
one or both of the polarizations from this path. \ This is equivalent to
blocking one or both of the LO\ beams. \ Ten percent of the light from path
2 is superposed with the down-conversion pump from path 1 at BS 2. \ The LO
beams are thus subject to the same spatial and spectral filtering as the
down-conversion and are separated by their polarizations at the PBS.

To demonstrate the interference effect, we adjusted the ND\ filters so that
the coincidence rate from the down-conversion path was equal to the
coincidence rate from the laser path. \ The polarizer was set to 45$^{\circ }
$, to transmit both horizontally and vertically polarized LOs. \ The singles
rates from the down-conversion path alone were 770 s$^{-1}$ and 470 s$^{-1}$
for detectors A and B respectively, and the coincidence rate was $\left(
3.3\pm 0.3\right) $ s$^{-1}$ (the ambient background rates of roughly 6000 s$%
^{-1}$ for detector A and 8000 s$^{-1}$ for detector B have been subtracted
from the singles rates, but no background subtraction is performed for the
coincidences). \ The singles rates from the LO\ paths were 11800 s$^{-1}$
and 53800 s$^{-1}$ for detectors A and B respectively, and the coincidence
rate from this path is $\left( 3.3\pm 0.3\right) $ s$^{-1}$. \ The LO
intensities need to be much higher than the down-conversion intensities to
achieve the same rate of coincidences because the photons in the LO\ beams
are uncorrelated. \ As the trombone arm was moved to change the optical
delay, we observed a modulation in the coincidence rate (Fig. 3.). \ The
visibility of these fringes is $\left( 48\pm 1\right) \%$, and when we
correct for our background (``accidental'') coincidences the visibility is
57\%. \ In theory, this visibility asymptotically approaches 100\% in the
very weak beam limit for equal coincidence rates from the down-conversion
and the LO paths. \ At the peak of this fringe pattern, the total rate of
photon pair production is greater than the sum of the rates from the
independent paths. \ This is an enhancement of the rate of photon pair
production. \ At the valley of the fringe pattern, the rate of the photon
pair production is similarly suppressed. \ The spacing of the fringes is at
the period for the 405-nm pump laser (approximately 1.3 fs/fringe).

The interference in the coincidence rate has been described as an
enhancement or a suppression in the rate of photon-pair production and
because of this there should be an accompanying change in the \textit{%
intensity} of the light reaching the detectors, and not merely the
coincidence rate. \ Fig. 4. shows four sets of singles-rate data for
detector A corresponding to four different polarizer settings. \ Recall that
the light was incident upon the polarizer at 45$^{\circ }$, so when the
polarizer is set to 45$^{\circ }$, both of the LO\ beams are free to pass. \
When the polarizer is set to 0$^{\circ }$ or 90$^{\circ }$, one of the LO
beams is blocked, and when the polarizer is set to -45$^{\circ }$ both of
the LO\ beams are blocked. \ The left hand side of Fig. 4 shows the data for
the two orthogonal diagonal settings of the polarizer, -45$^{\circ }$ (top)
and 45$^{\circ }$ (bottom); the right hand side shows the data for the two
orthogonal rectilinear settings, 0$^{\circ }$ (top) and 90$^{\circ }$
(bottom). \ For the 45$^{\circ }$ data, the singles rate at detector A\
shows fringes with a visibility of about $0.7\%$. \ This visibility is
roughly 100 times smaller than the corresponding visibility in the
coincidence rate because only about 1\% of detected photons are members of a
pair, due to the classical nature of our LO\ beams. \ The fringe spacing in
the singles rate corresponds to that of the pump laser light at 405 nm even
though it is the 810-nm intensity that is being monitored. \ By examining
the other three polarizer settings (-45$^{\circ }$, 0$^{\circ }$, and 90$%
^{\circ }$), it is apparent that in order to observe fringes in the singles
rate, both LO\ paths must be open. \ This demonstrates that we are observing
a nonlinear effect of one polarization mode on another, at the single-photon
level.

When destructive interference reduces the intensity of the beams reaching
the detectors, energy conservation dictates that all incident laser photon
pairs must be undergoing sum frequenc generation. \ To explicitly verify
that photon pairs are actually removed from the LO beams, a simple extension
was performed. \ In the presence of the strong classical pump, it would be
impossible to observe the upconverted photon directly, so we measure the
reduction in the coincidence rate relative to the coincidence rate from the
LOs alone. \ In order to maximize the effect, the coincidence rates from the
LO path and the down-conversion path were set to $\left( 38.2\pm 0.7\right) $
s$^{-1}$ and $\left( 1.2\pm 0.2\right) $ s$^{-1}$, respectively. \ The
coincidence rate was again recorded as a function of the optical delay and
is shown by the filled circles in Fig. 5. \ A sinusoidal fit to the data is
shown as a heavy black line, and has a fringe visibility of $\left( 19.0\pm
0.5\right) \%$. \ The coincidence rate from the LO\ paths alone was measured
before and after the experiment was performed and is shown as a horizontal
dashed line, as well as an open square with error bars indicated. \ For
delay positions where the solid black line drops below the dashed line, the
photon pair production rate drops below the rate from the LO rate alone. \
This reduction in the pairs is due to photon pairs being removed from the LO
beams undergoing sum-frequency generation. \ From the fringe visibility, we
can infer that at least $\left( 15.7\pm 1.7\right) \%$ of the photon pairs
from the local oscillator were converted into the second harmonic. \ This
corresponds roughly to a few tenths of a percent of the photons overall.

We have demonstrated a quantum interference effect which is an effective
nonlinearity at the single-photon level. \ We have shown that pairs of
photons may be removed from a pair of LO\ beams, and that their removal
results in the reduction of the number of photon pairs reaching a pair of
detectors. \ This effect is accompanied by a corresponding change in the
intensity of the beams. \ The effect studied in this work is closely
analogous to second-harmonic generation in traditional nonlinear optical
materials, but is enhanced by the simultaneous presence of a strong
classical pump with appropriately chosen phase. \ For a different choice of
phase, it should be possible to observe an effect analogous to cross-phase
modulation. \ Such a conditional phase shift may contribute to the
realization of a controlled-phase gate for photons. \ These nonlinearities
may also be useful for the problem of distinguishing the four
maximally-entangled Bell-states that are ubiquitous in quantum optics. \ It
has been shown that it is impossible to distinguish between all four states
with only linear optics \cite{bell-state}. \ Work on unconditional
teleportation \cite{shih} has been limited to efficiencies near 10$^{-12}$.
\ It may be possible to build a scheme centered around an effect like the
one demonstrated in this work that would be capable of distinguishing all
four states. \ Overall, effects such as this hold great promise for
extending the field of nonlinear optics into the quantum domain.

We are grateful for the financial support of NSERC, CFI,\ and Photonics
Research Ontario. \ One of us, K.R., would like to thank the Walter C.
Sumner Foundation for financial support.

\bigskip

Fig. 1. A simplified cartoon of our experiment. \ a)\ Pairs of weak coherent
states, or local oscillator (LO) beams (shown by dashed lines) are
overlapped with the pair of down-converted photon beams. \ A coincidence
count is registered either if b) a down-conversion event occurs, \textit{or}
if c) a pair of laser photons reaches the detectors (SPCMs). \ The e$%
^{i\varphi }$ in figure c)\ represents a controllable relative phase between
the two Feynman paths that lead to a coincidence.

Fig. 2. A schematic for the setup of the experiment. \ BS 1 and BS 2 are
90/10 (T/R) beamsplitters; SHG consists of 2 lenses and a BBO\ nonlinear
crystal for type-I second harmonic generation; BG 39 is a coloured glass
filter; ND\ is a set of neutral density filters; $\lambda $/2 is a
zero-order half-wave plate; DC\ is a type-II down-conversion crystal; PH is
a 25-$\mu $m diameter circular pinhole; I.F. is 10-nm-bandwidth interference
filter, PBS is a polarizing beam splitter; and Det. A and Det. B\ are
single-photon counting modules. \ The thinner solid line shows the beam path
of the 810-nm light, and the heavier solid line shows the path of the 405-nm
light.

Fig. 3. The coincidence rate as a function of the delay time. \ The
interference is a phase-dependent enhancement or suppression of the photon
pairs emitted from the crystal. \ The visibility of these fringes is $\left(
47.4\pm 0.9\right) \%$, and once corrected for background the visibility is
57\%.

Fig. 4. The singles rate at detector A versus the delay for 4 different
polarizer angle settings. \ The left-hand data sets are for the polarizer
settings of $\pm 45^{\circ }$. \ The right-hand column is for the polarizer
settings of 0$^{\circ }$ and 90$^{\circ }$. \ The fringes are apparent only
for the +45$^{\circ }$ polarizer setting, and have a visibility of $0.7\%$.
\ These four data sets show that both horizontally \textit{and}
vertically-polarized photons must be present for the effect to occur.

Fig. 5. The coincidence rate versus the delay for a case where the rates
from the different Feynman paths are severely imbalanced. \ This
demonstrates that some photon pairs from the LO\ beams are being
upconverted. \ The average value of the coincidence rate from the laser path
alone is represented by a hollow square and the dashed horizontal line. \
The solid curve is a sinusoidal fit to our data. \ It is apparent for
certain delay (phase) settings the coincidence rate drops below the rate of
coincidences from the LO\ paths alone. \ Based on the magnitude of this
drop, we conclude that at least $\left( 15.7\pm 1.7\right) $\bigskip $\%$ of
the photon pairs from the laser are upconverted.

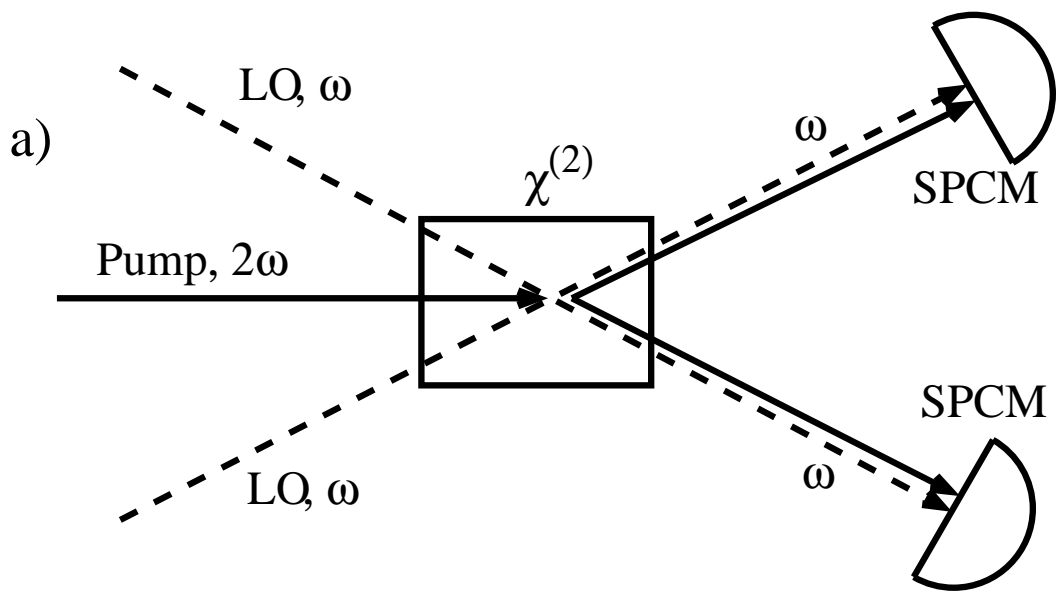
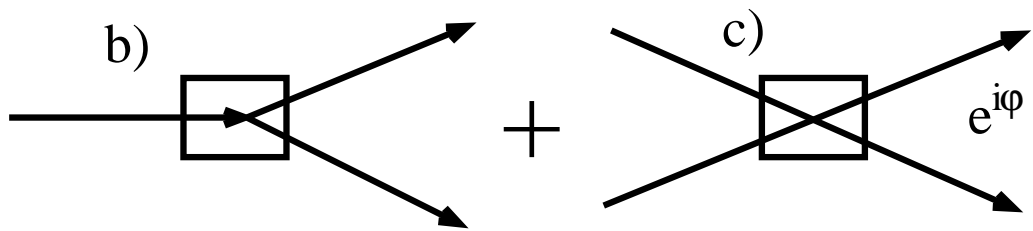
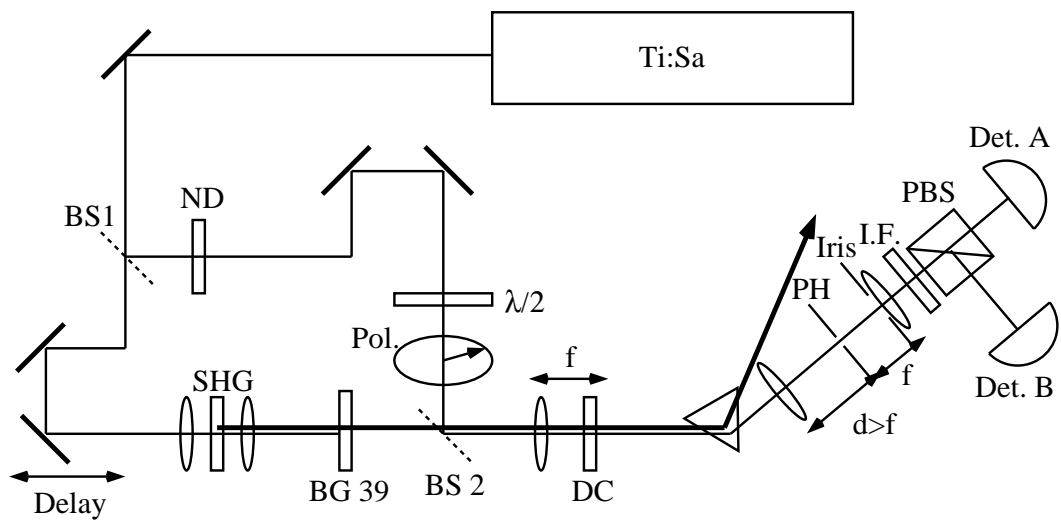
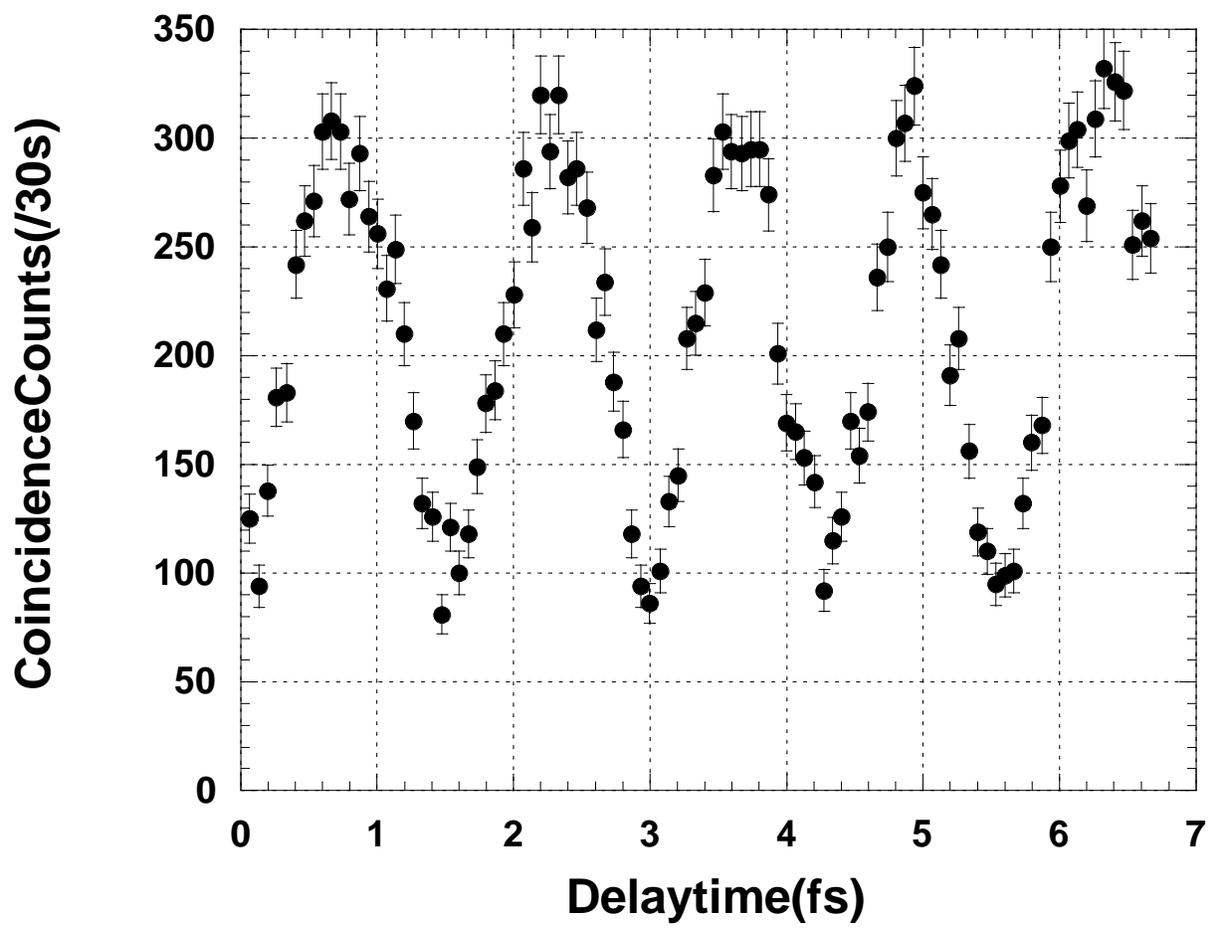
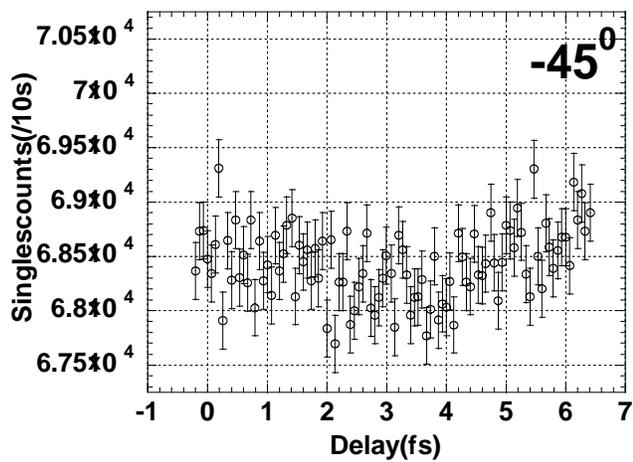
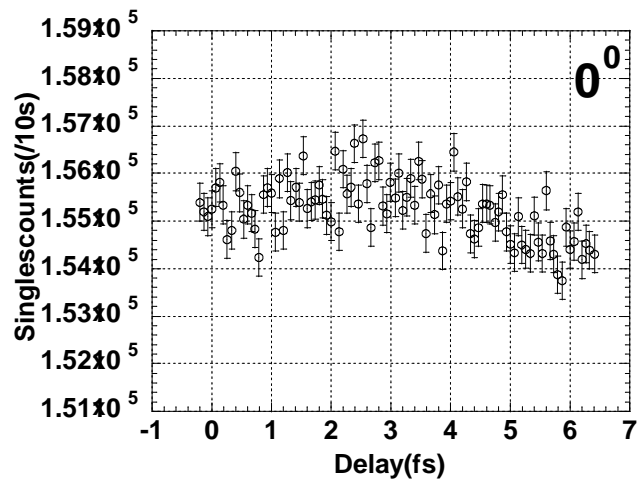
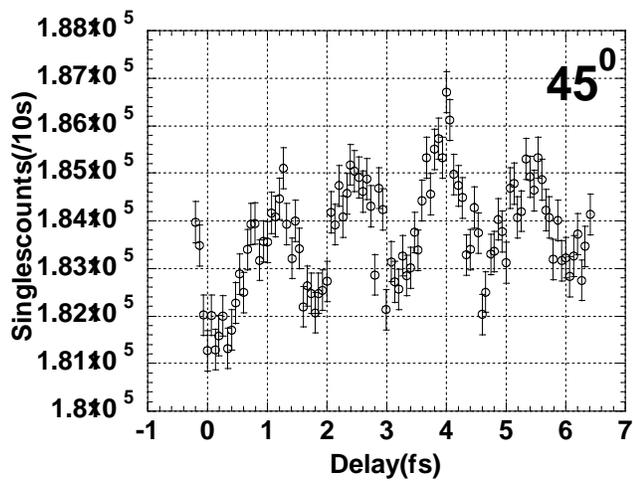
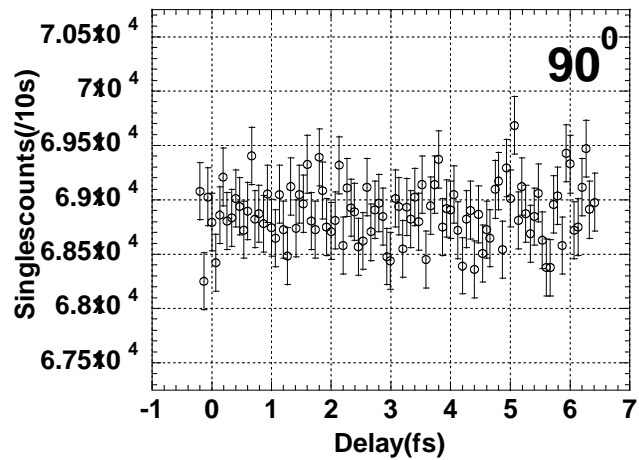
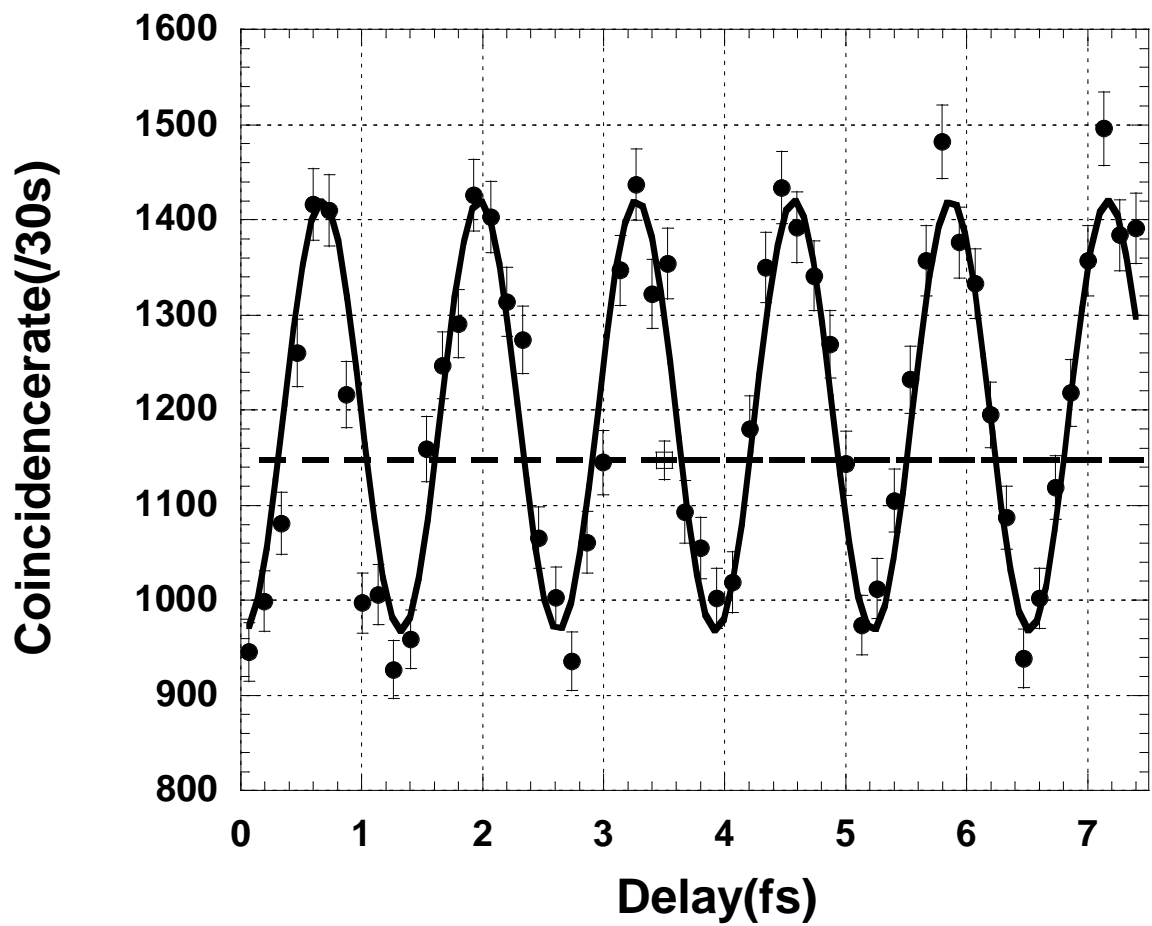
\end{document}